**Study of RF Sputtered Antimony Alloyed Bismuth Vanadium Oxide (Sb:BiVO$_4$) Thin Films for Enhanced Photoelectrochemical (PEC) performance from Bandgap Modulation to Thickness Optimization.**

*Tilak Poudel, Yanfa Yan, Xunming Deng*

**Abstract**

Monoclinic scheelite bismuth vanadate (BiVO$_4$) is a promising photoanode for water splitting yet the PEC performance is limited due to its relatively higher (2.4 eV) band gap. Here, we successfully decreased its the band gap to 1.72 eV by controlled antimony alloying. Low bandgap antimony alloyed bismuth vanadium oxide (Sb:BiVO$_4$) thin film was prepared by RF sputtering of high purity homemade target, fabricated by solid-state reaction using a mixture of Sb$_2$O$_3$, Bi$_2$O$_3$, and V$_2$O$_5$ powders with desired stoichiometric ratios. Several growth parameters, powder crystallography, post-deposition effects, and surface treatments, thickness dependence, effect of electrolytes on photocorrosion were studied along with its optical and electrochemical characterization. We discovered that Sb:BiVO$_4$ is a direct band gap material in the visible light range (1.72 eV) and a valence band position suitable for driving water oxidation reaction under illumination. Furthermore, hole diffusion length is increased with antimony alloying and achieved optimum thickness of 400 nm for higher photocurrent. The controllably prepared Sb:BiVO$_4$ particles are having the sizes of 10-15 nm in room temperature deposition and can be grown up to 0.5 microns under air annealing.

**Introduction**

The increasing global awareness concerning carbon emissions and the exploitation of fossil fuel reserves motivates the development of technology based on alternative energy sources [1]. With 173,000 TW of solar energy striking the surface of the earth at any given moment, the challenge is to convert sunlight into useable form of energy. Solar Photovoltaic (PV) cells normally generate electricity using sunlight but due to the variability of daily solar irradiance, harvesting sunlight into a storable chemical energy has been considered an essentially sustainable pathway to mitigate world's energy crisis [2] [3] [4]. This chemical energy could be in the form of hydrogen, which has highest energy density per mass of 142 MJ/Kg [5].

Hydrogen has the potential to be a sustainable carbon-free fuel, but it is not available as a primary source in nature [6]. Nevertheless, it can be produced from renewable sources (water and sunlight) and converted into electricity at relatively high efficiency with environment friendly end products. The direct photoelectrolysis of water was first achieved by Fujishima and Honda in 1972 with TiO$_2$, a large band gap semiconductor photoelectrode [7]. However, owing to its larger band gap only a limited portion of solar spectrum can be used. Most of the binary oxide semiconductors (e.g. Fe$_2$O$_3$, WO$_3$, TiO$_2$, SnO$_2$, ZnO, CoO$_x$) available for water splitting have a large energy band gap or not too many ternary oxides have the band edge potentials suitable for oxygen and hydrogen evolution [8] [9].

Bismuth vanadate (BiVO$_4$) semiconductor materials have been extensively studied as a promising photoanodes for photoelectrochemical water splitting due to its relatively narrow band gap of 2.4 eV, favorable band structure, earth abundance, low toxicity, chemical stability, long hole diffusion length, and substantial visible light absorption [10] [11]. Many synthesis strategies including precipitation reactions [12] [13], hydrothermal synthesis [14] [15] [16], sol-gel methods [17] [18] have been reported for the preparation of BiVO$_4$ powders. As a drawback, bismuth vanadate semiconductor has a conduction band edge positioned at an energy level inferior of the reversible hydrogen potential. The major performance bottleneck is poor separation of the photoexcited electron–hole pairs due to the extremely low carrier mobility ($\sim 10^{-2}$ cm$^2$ V$^{-1}$ s$^{-1}$), which results in significant carrier recombination losses [19] [20] [21]. Consequently, bismuth-based devices need an external bias voltage to promote water photoreduction. For the efficient charge transport in the system, the optimized thickness of the BiVO$_4$ photoanodes should match its charge carrier diffusion length (L$_d$ = 70-100 nm) [22] [23] [11]. Nevertheless, the carrier diffusion length also changes with antimony incorporation.

Bulk and surface recombination losses of photogenerated charged carriers play a vital role in determining the efficiency of oxide photoanodes such as BiVO$_4$. Low mobility and short minority charge carrier lifetime lead to poor collection of holes from the bulk of the material to the surface of the material [21]. The short diffusion length has been addressed by fabricating nanocrystalline films in which the characteristic dimensions are comparable with diffusion length, so that holes from the n-type photoanodes gain significant chances of reaching to the surface. Recent optimization on surface morphology [24] and elemental doping (such as W, Mo, Ni, Nb etc.) into BiVO$_4$ bulk have resulted in substantial improvements in the photocurrent response in concentrated photo illumination [25] [26] [20] [1] [27] [28] [29] thereby obtaining higher charge separation efficiency and reducing surface recombination. In this report, we synthesized reduced bandgap BiVO$_4$ thin film by alloying with antimony in Sb:V = 1:10 ratio.

**Experimental Methods**

1) **Target synthesis: Preparation of Sb:BiVO4 precursor powder**

Precursor powder was synthesized by a high-temperature solid-state reaction method by mixing bismuth oxide (Bi$_2$O$_3$), Vanadium oxide (V$_2$O$_5$) keeping the metals ration Bi:V = 1:1 and added Sb$_2$O$_3$ such that ratio of Vanadium and Antimony becomes Sb:V =1:10. The mixture was homogenized using a rolling mixer and as-obtained yellowish precursor powder was transferred to a fused-silica crucible which was placed into an air- ambient electric muffle furnace at 840 ºC for 140 hrs. A fused-silica plate was placed over the crucible to mitigate the loss of any volatile components (particularly Bi vapor) without creating a gas-tight seal, thus allowing excess oxygen to be present during the annealing process. The assembly was brought up to approximately 840 ºC over a period of 4 hours and then held at this temperature for another 140 hours. The furnace was then deactivated and allowed to cool naturally in the closed state. The resulting powder was strong solid chunk and was mild yellow in appearance. This solid material was then crushed and grinded thoroughly using agate mortar and pestle until it became a fine microparticle.

Approximately 20 g of the annealed powder was used to fabricate a target which could be used for RF sputtering depositions. The powder was loaded into a stainless-steel target cup with a 2″ diameter cavity and pressed at room temperature using hydraulic press with an applied force of 12 tons for 20 minutes to get the high purity target ready for installation.

## 2) Thin Film Deposition: Radio Frequency (RF) Magnetron Sputtering

The high-purity homemade target was loaded into a custom-built sputtering chamber. Depositions were performed on Fluorine-doped tin oxide coated glass substrates (FTO-Tec 15 Pilkington). Substrates were cleaned ultrasonically with DI water, acetone, isopropanol, and ethanol before loading into the RF sputtering chamber. The substrate was covered by a strip on one side to prevent film deposition underneath so that electrical contacts could be made for the electrical and photoelectrochemical measurements.

The Sb:BiVO$_4$ thin films were prepared by Radio Frequency sputtering in Argon/Oxygen plasma environment. RF sputtering powers typically ranging from 40 – 70 W, and chamber pressures of 10 mTorr sustained by supplying argon gases with the flow rate of 30 sccm. The thickness of the obtained film was varied from 80 nm to 1 micron.

## 3) Characterization and measurement

The atomic ratios of metals in the precursor powder were determined using energy dispersive x-ray spectroscopy with Rigaku Cu Kα radiation. The surface morphology and bulk elemental composition of the thin films were characterized using a Hitachi scanning electron microscope (SEM) with in-built energy dispersive spectroscopy (EDS) attachment. EDS measurements for elemental analysis were taken of regions approximately 500 μm x 500 μm in area. Film thicknesses were measured using a DEKTAK profilometer to determine the step height at two locations namely at a tape-masked center and holder frame-masked edges to observe the thickness variation across the film. Bulk crystalline structure of the thin films was characterized using a Rigaku X-ray diffractometer using coupled 2θ Bragg-Brentano mode and a copper X-ray source (Kα Cu =1.54 Å). Phase assignments were made based on the Joint Committee on Power Diffraction Standards (JCPDS) database.

The optical absorbance spectrum was measured by a PerkinElmer lambda 1050 UV-vis-NIR spectrophotometer. The transmittance and reflectance of the samples were measured by optical spectrometer using 300 nm – 1500 nm wavelength and the band gaps were calculated using the following relation.

$$\alpha = \frac{1}{t} \ln \left[ \frac{(1-R)^2}{T} \right] \quad (1)$$

Where α is the absorption coefficient, t is thickness, and R and T are reflection and transmission respectively. Now, if we plot (αhν)$^n$ vs. hν, the we can get a straight line, the intercept of which gives us the band-gap value.
n= 2 for direct; n=1/2 for indirect transition.

## 4) Photoelectrochemical (PEC) measurements

Photoelectrochemical measurements of this oxide photoanodes were conducted using a three-electrode cell configuration in various electrolytes with Ag/AgCl reference electrode and a platinum coil as a counter electrode. The major advantage of integrated PEC is that solar energy capture, conversion, and storage are combined in a single integrated system. Films were tested in various electrolyte solutions such as 1M KOH, 1 M NaOH, NaSO$_3$ + KH$_2$PO$_4$, and H$_2$SO$_4$ to observe the photocatalytic performance and resulted chemical corrosion. $H_2O_2$ was also added as a sacrificial agent to improve oxygen evolution reactions in certain cases. Voltammetry experiments were performed using a computer-controlled potentiostat. For PEC measurements, the electrode was illuminated from the front side (through the electrode) and back-side (from the glass side) as well using 15 W xenon lamp with AM 1.5. The light intensity was calibrated using a silicon diode. The illumination area was typically around 1 $cm^2$ or less.

The photoelectrochemical properties were investigated in three electrode configurations using Sb:$BiVO_4$ films as working electrode. Sb:$BiVO_4$ electrodes were prepared by cutting large 3″×3″ thin films deposited on Tec-15 substrates into smaller 0.75″×1.5″ rectangular shapes. Thin film was then covered by a non-conducting epoxy resin leaving behind a smaller area (typically less than 1 $cm^2$) for light exposure. The film on one side of these electrodes was then etched away by gentle mechanical scratch exposing underlying conducting layer of the substrate and coated with thin indium metal layer for better electrical contact. The PEC characterization was carried out using Voltalab potentiostat, in a three-electrode configuration in a quartz-windowed cell partially filled with electrolyte solution. The recorded potential versus Ag/AgCl ($E_{Ag/AgCl}$) in this work was converted into potential against reversible hydrogen electrode (RHE) using the Nernst equation (2) given below:

$$E_{RHE} = E_{Ag/AgCl} + 0.059 \times pH + 0.1976\ V \qquad (2)$$

The system was purged with nitrogen for 30 mins in order to remove possible oxygen dissolved in the electrolyte. Photoelectrochemical response was recorded on both the forward bias and reverse bias potential under illumination. The illumination source was 300 W Xe lamp calibrated and equipped with AM 1.5 filter. The light intensity of 100 mW/$cm^2$ was adjusted and calibrated using silicon photodiode.

**Results and discussion**

### A) Crystal Structure and Surface Morphology

The XRD patterns of annealed $BiVO_4$ films with different elemental doping concentrations have been studied. X-ray diffraction patterns in Figure 1 shows that upon 10% antimony alloying Bragg peaks of $BiVO_4$ (011) disappeared and intensity of peak (112) at $27^0$ increases. The increase in intensity suggests a preferential growth of 112 orientation, which may be arisen from Sb affecting film nucleation during sputtering process. This was not expected in case of substitutional doping or replacement of $V^{5+}$ ions with $Sb^{3+}$ ions since $Sb^{3+}$ ions (0.76 Å) are bigger than $V^{5+}$ ions (0.54 Å). If $Sb^{3+}$ is present in the interstitials as a dopant or replaces $Bi^{3+}$ (ionic radius of 1.17 Å), there would have been decrease in lattice parameter, this trend can be seen in XRD pattern for $SbVO_4$ comparing with $BiVO_4$ (supplementary information), where the peak has shifted to higher 2theta value [21]. When doping concentration is more into alloying label, the crystal structure is not retained, and thence the presence of secondary phases is revealed. We suspect the presence of both the $V^{3+}$ and $V^{5+}$ ions in the $BiVO_4$ crystal lattice and a fraction of $Sb^{5+}$ ions (0.60 Å) replaces $V^{3+}$ ions (0.64 Å) during alloy sintering process.

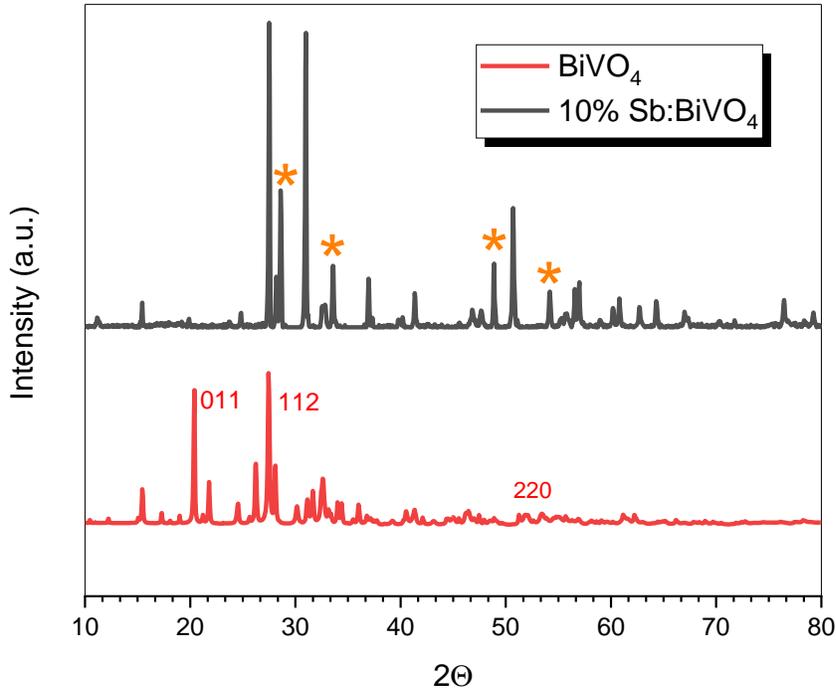

*Figure 1: XRD patterns of undoped BiVO₄ samples (red curve) and 10% antimony alloyed BiVO₄ samples. The peaks indicated by "*" originated from underlying FTO substrate.*

The uniform film of thickness ranging from 80 nm to 1 µm was obtained by sputtering antimony alloyed BiVO$_4$ target in Argon plasma (a total flow of 30 sccm) using RF power of 50 - 70 W. The synthesized target and as deposited sputtered film are shown in Figure 2. The deposition rate was 1.5 – 2.0 nm per minute.

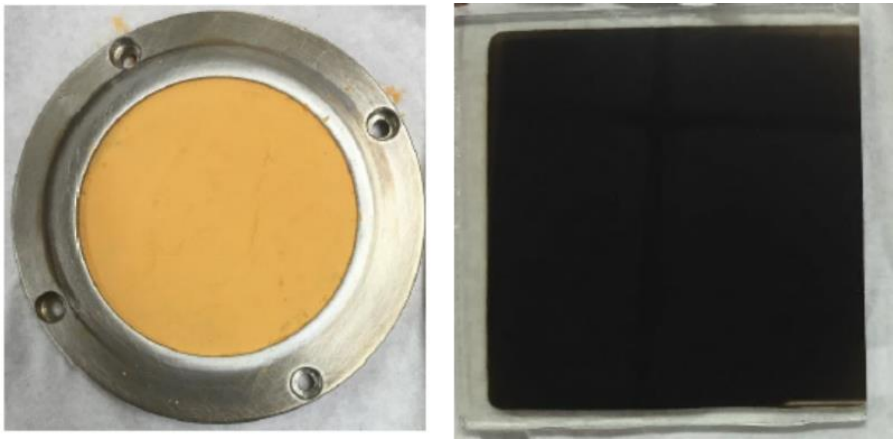

*Figure 2: Photographs of Sb alloyed BiVO4 homemade target and RF sputtered thin film sample of thickness of 800 nm.*

The surface morphology of the Sb:BiVO4 films is illustrated by the SEM pictures in Figure 3. The room temperature deposited film shows evidence of evidence of uniform crystal growth with average grain sizes of 50 nm. EDX studies confirmed the concentration level of antimony into the bulk. The films appear to be non-porous, and crystal size considerably grows bigger upon subsequent annealing. It can be seen that the surface of the films consists of well-formed crystallites with lateral dimensions of the order of 0.1-0.5 microns after annealing at 200 ºC for 60 minutes. The relatively smooth morphology and compactness compared with nanostructured films is convenient for this study since it avoids the complications associated with three-dimensional development of the space charge region.

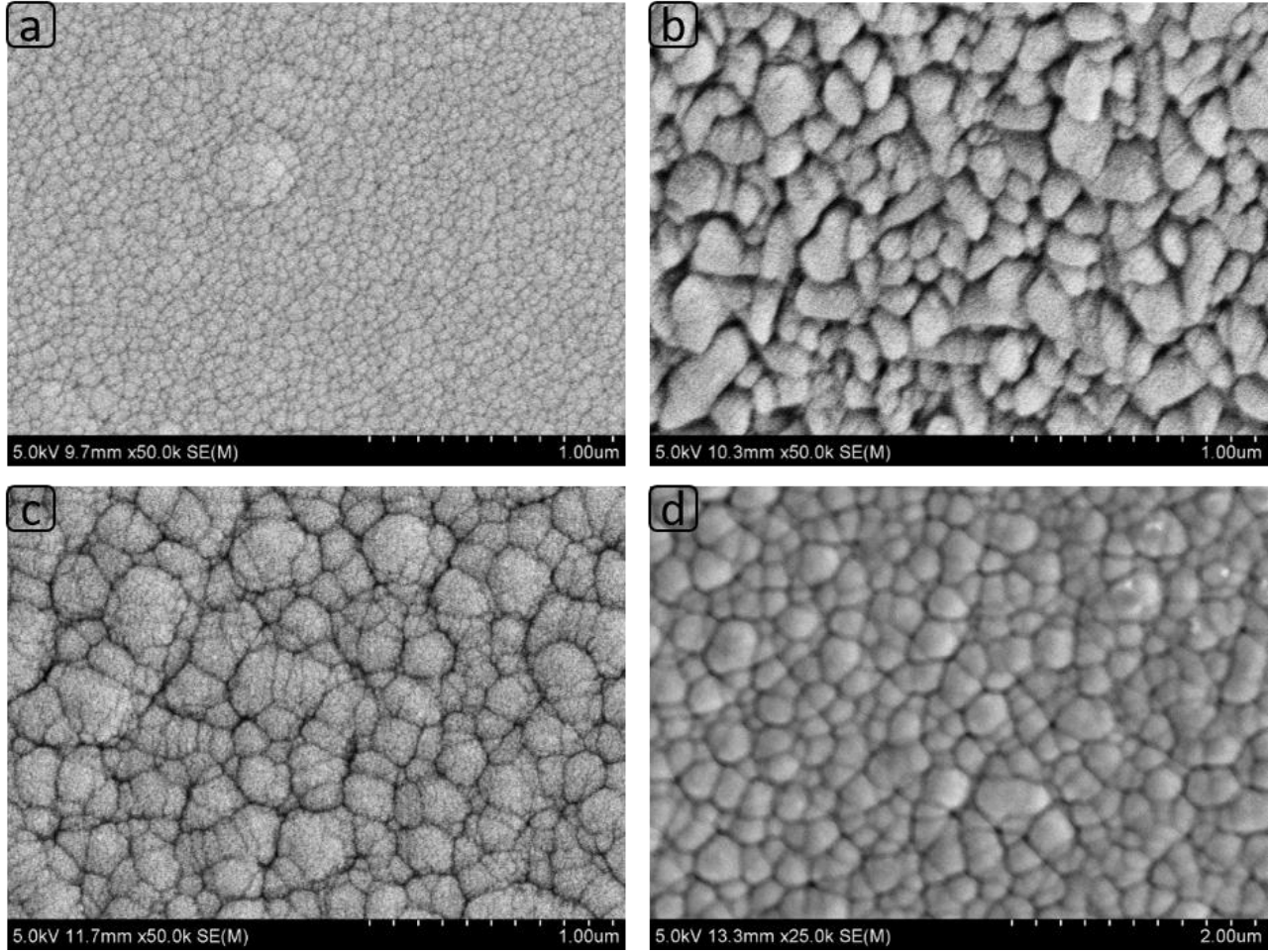

*Figure 3: SEM micrographs a) as deposited Sb alloyed BVO at room temperature; b) annealed at 100 ºC; c) and d) annealed at 200 ºC for 60 minutes.*

### B) Optical Properties

The tauc plot for direct band gap estimation of 10% antimony alloyed $BiVO_4$ (Sb:$BiVO_4$) thin films is shown in Figure 4 below. As in literature [1] [30] [31], undoped $BiVO_4$ has a direct band gap of 2.4 eV. Researchers suggest that substitutional doping is unique method to reduce its intrinsic band gap [32] [33]. Our research indicates that with the incorporation of antimony in the bulk of the film, the band gap can be easily manipulated. Unlike the doping, antimony alloying does not provide linear relationship between alloying concentration level and band gap of the material. We found that with higher alloying level of Sb to V as 1:1 and more lowers the band gap by 0.13 eV but it also produces secondary phases. For instance, band gap of 2.27 eV for

$BiSb_{0.5}V_{0.5}O_4$ as shown in supplementary information. But with lower percentage of antimony alloying into the film, we obtained different results with significant reduction in band gap. 10% antimony alloyed (Sb:V = 0.1:1) BiVO4 exhibits significantly lower band gap of 1.72 eV and corresponding enhanced photo absorption as shown in Figure 4 and Figure 5.

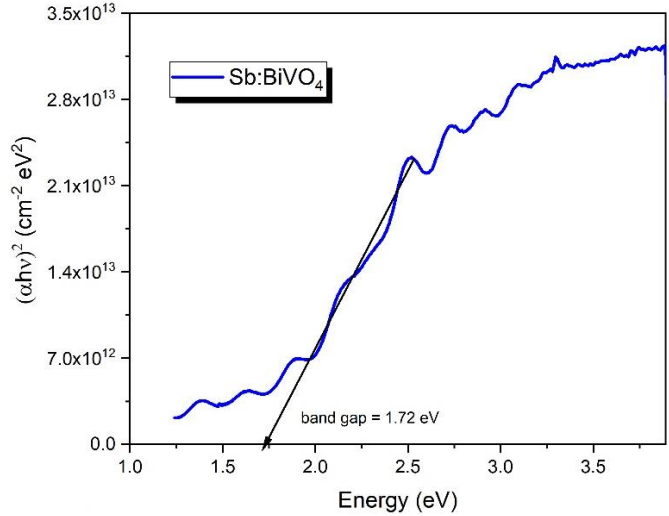

*Figure 4: Tauc Plots for direct band gap estimation a) 0.1 Sb alloyed $BiVO_4$ annealed at 200 °C*

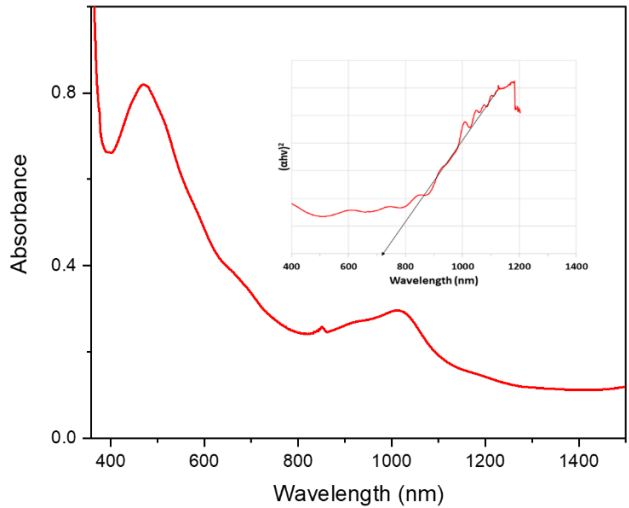

*Figure 5: UV-visible absorption spectra of 400 nm thick Sb:BiVO4 thin film. (Inset: direct band gap estimation after annealing at 200 °C for 60 minutes)*

### C) Photoelectrochemical testing of sputtered Sb:BiVO4 photoanodes

The photocurrent-voltage characteristic of a photoelectrochemical cell for solar hydrogen production via water splitting, using 10 % antimony alloyed $BiVO_4$ as photoanode, was obtained. Photoelectrochemical characteristics of the cell were investigated by three electrode system. Radio

Frequency sputtered Sb:BiVO$_4$ thin films displayed significantly improve on its photochemical activities under forward bias compared to RF sputtered pristine BiVO$_4$ (not shown here). The photocurrent obtained for rf sputtered Sb:BiVO$_4$ electrodes mostly depends on its thickness. In general, photo absorption increases with thickness of the absorber layer up to an optimized limiting thickness. Thinner films suffer from large dark current at higher bias region, thicker films show smaller photocurrent at higher bias. Roughly 400 nm thick Sb:BiVO$_4$ exhibits higher photocurrent, which is evident that hole transport length increases with antimony alloying compared to 200 nm thick pristine bismuth vanadate [34]. Optimum thickness is 400 nm where film displays significantly higher photocurrent at lower bias region as shown in Figure 6b. However, further increases in thickness does not necessary improve photocurrent as Figure 6a showed that hole transfer length should be in the order of film thickness to gain maximum photocurrent.

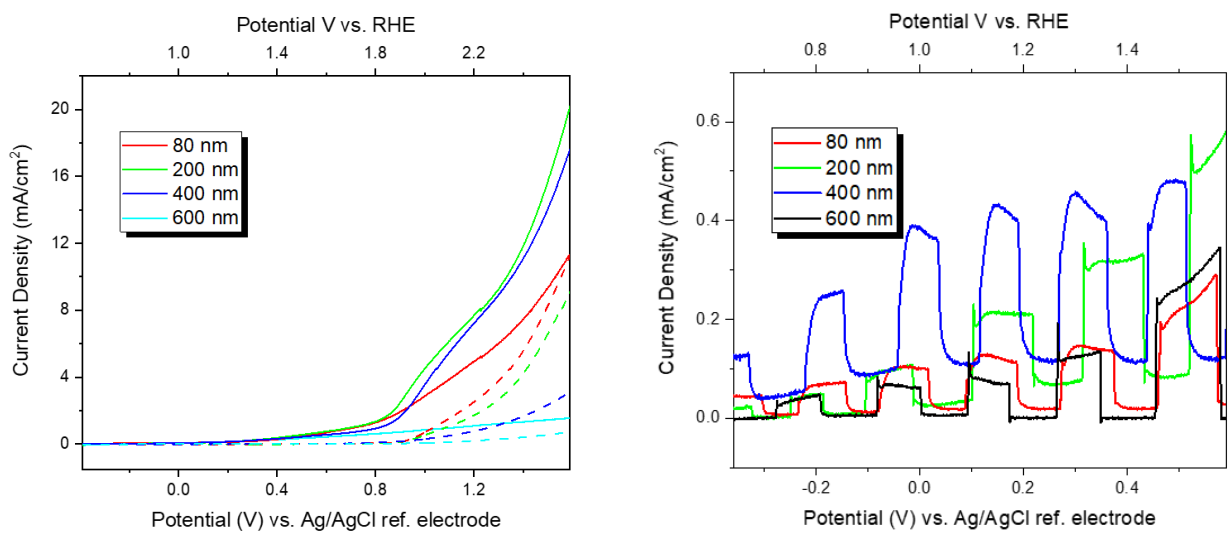

Figure 6: JV curve for 80-600 nm thick sputtered Sb:BiVO$_4$ on FTO in pH 13.7 KOH solution a) under dark and illumination. b) chopped light

The dark current increases in higher bias region and it is more pronounced in thinner films than the thicker films as shown in Figure 6, dotted red lines for 80 nm thin film increases sharply after 0.9 V against Ag/AgCl but the steepness decreases as we go to thicker films and there is not significant dark current for 600 nm thick film as in dotted sky-blue lines. Clearly this means that accumulation of positive surface charge (i.e. holes) associated with Fermi level pinning resulted from crystallographic defects and an indication of uncompensated series resistance losses at the surface [35]. Furthermore, the spikes and transient current overshoot on the photocurrent plots (as in Figure 6b) is due to the build-up of positive charges that are queuing to take part in the OER. These trapped holes at the surface appears to be associated with sluggish hole transfer kinetics, which in theory can be overcome by using suitable catalyst for OER or by adding hole scavenger into the electrolyte as shown in Figure 7. In addition, electrodes tested in pH 13.7 electrolyte solution suffer from severe chemical corrosion. The reason for chemical corrosion is unknown at this point. So, we performed other tests on lower pH electrolyte (pH 7.0 buffer) to conform that the corrosion is typically chemical in nature but not due to photocorrosion. Surface treatments or extra layer deposition on the top would open the pathways for applications of Sb:BiVO$_4$ in stacked multijunction electrodes. At the same time, recombination losses in the bulk of the semiconductor electrode can be minimized by growing nanostructured thin films or incorporated nanoporosity [36] [20].

It can be seen form **Error! Reference source not found.**(supplementary information) that the cathodic shift of photocurrent onset potential to -0.35 V for 10% antimony alloyed bismuth vanadate. There are several possible reasons for this behavior, including widening of the space charge and/or reduction in surface recombination as well as the kinetics of the multi-electron transfer process.

The PEC performance of antimony alloyed bismuth vanadate was tested under lower pH buffer solution to study surface kinetics without suffering from electrode corrosion. Even though the overall photocurrent is low in neutral electrolyte as compared to higher pH solution, we found that with the addition of $H_2O_2$ hole scavenger, the photocurrent increases significantly as shown in Figure 7 below.

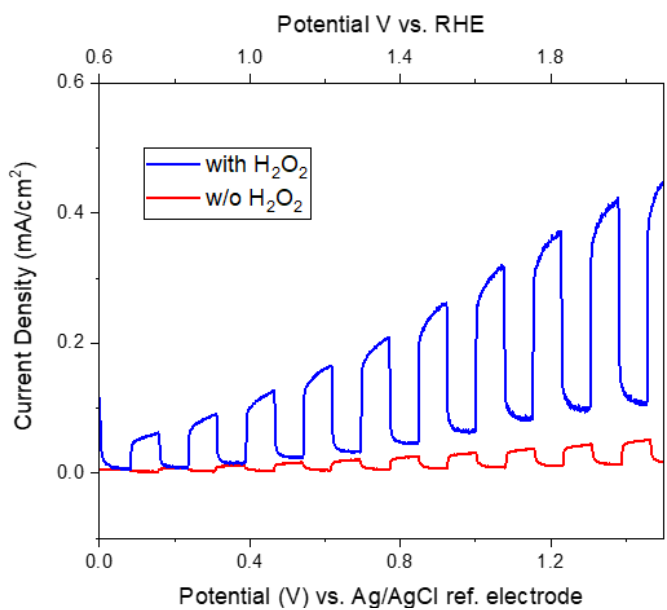

*Figure 7: Photocurrent response of Sb:BiVO4 electrode with and without hole scavenger (3% $H_2O_2$) in pH 7 buffer solution.*

In addition to the relationship between doping concentration, thickness dependency, and PEC performance from sputtered Sb:BiVO$_4$ films, we also explored differences between front illumination and back illumination as shown in Figure 8. A variety of reports on solution-based BiVO4 thin films have found superior photocurrent under back-side illumination because of significantly lower charge separation efficiency [37] but can be increased by extra treatment or intentional doping [38]. But in contrast, we observed the net photocurrent under front-side and back-side illumination is almost uniform for 400 nm thick film in pH 7.0 buffer electrolyte. Our study finds that the maximum photocurrent obtained with 400 nm thick samples may result from the space-charge region widening with antimony alloying in BiVO$_4$ lattice. Literature suggests the highest photovoltage is obtained for 100 nm thick pristine BiVO4 because of its matching hole diffusion length [39]. When the film thickness exceeds the hole diffusion length, holes generated deeper in the bulk could not transfer to the surface to involve in oxidation reaction and gets trapped or recombine with majority electron. However, when hole acceptors like $H_2O_2$ with fast oxidation kinetics were present in electrolyte, the higher photocurrent generates suggesting the presence of considerable surface states causing rapid surface recombination [40]. For the thickest film of 1 micron, neither front nor back-side illumination is effective since obtained photocurrents were next to negligible.

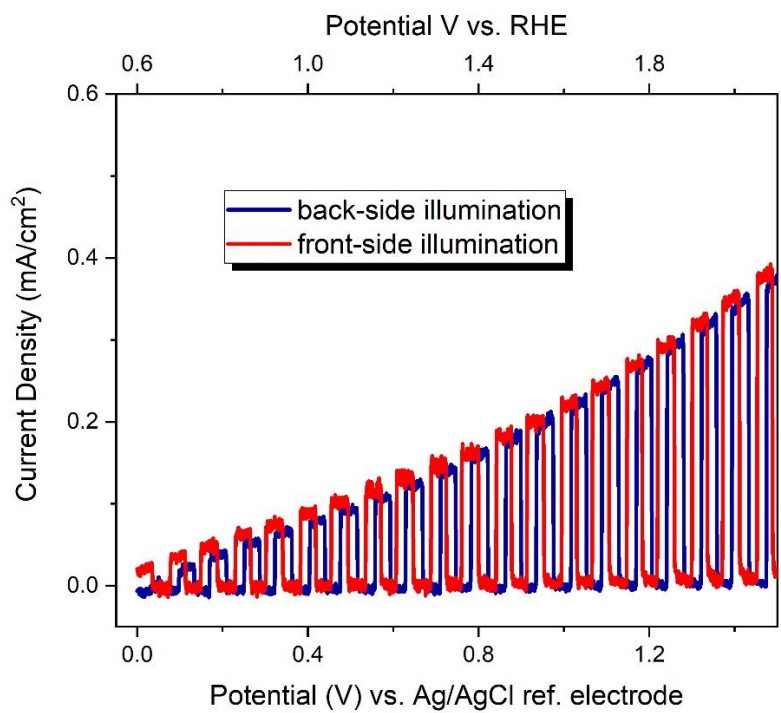

*Figure 8: J–V (current density vs voltage) curves of 400 nm thick Sb:BiVO4 photoandoe. The samples were tested in a three-electrode configuration under front and back-side chopped illumination with an AM 1.5, 100 W/cm2 light source and in pH 7 buffer solution*

### D) Effect of Electrolytes

More recently, the effects of the electrolyte and its pH on the PEC performance and photocatalytic activities of pristine BiVO4 have been investigated [41] [42]. Researchers found that increasing the pH of the electrolyte solution from acidic to alkaline increased the band bending and hence electron-hole separation [43] [44], our observation aligns towards the same but highly alkaline electrolyte solution appears to etch away the material in longer exposure as shown in photographs below.

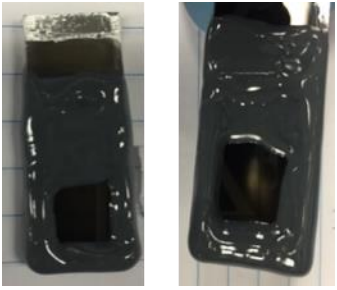

*Figure 9: Sb:BiVO4 electrodes before and after multiple testing in 1.0 M NaOH solution. A bright white spot on the center of the film appears to be chemical erosion of materials.*

We considered the visual degradation of thin film in 1.0 M NaOH solution as a form of chemical corrosion as dark current goes high under testing as FTO exposes directly to the electrolyte. Surface treatments and overlayers at the surface are generally implemented to reduce surface recombination [45] [46] and inhibits the chemical degradation of thin films. In our study, 2 nm gold underlayer deposited on FTO before thin film deposition helped to separate generated charged carriers and reduce the dark current and a very thin nickel overlayer (2-10 nm) deposited on top of Sb:BiVO$_4$ increases resistance against chemical corrosion but reduced photo absorption significantly in both cases as shown in supplementary information.

**Conclusions**

We synthesized high quality Sb:BiVO$_4$ thin films by reactive RF sputtering for the first time and explored the influence of the V/Sb ratio on structure, morphology, and PEC performances. Antimony alloying resulted in significant band gap reduction especially for 0.1:1 Sb:V ratio. Controlled antimony alloying changes valence band position thereby narrowing down the fundamental band gap of the material. 10% Sb alloyed BiVO$_4$ thin film showed significantly smaller band gap of 1.72 eV and relatively higher photocurrent density without any surface modification. Nonetheless, Sb:BiVO$_4$ displayed its applicability in commercial water splitting materials by modifying its surface to resist chemical corrosion.

We also discussed the influence of front illumination versus back illumination, finding that the influence of illumination side was film thickness dependent and might be limited by hole transport. In addition, we improved the resistance against photocorrosion by depositing thin layer of Nickel on the top of Sb:BiVO$_4$ thin film but at the expense of photocurrent. In the future, we will work on the further improvement of Sb:BiVO$_4$ thin films for PEC application such as improving the resistance against chemical corrosion and improving hole transfer by loading OER catalysts for the fundamental studies of bismuth based semiconducting materials.

# References

[1] T. W. Kim, Y. Ping, G. A. Galli and K.-S. Choi, "Simultaneous enhancements in photon absorption and charge transport of bismuth vanadate photoanodes for solar water splitting," *Nature Communications,* 2015.

[2] S. Y. Reece, J. A. Hamel, K. Sung, T. D. Jarvi, A. J. Esswein, J. J. H. Pijpers and D. G. Nocera, "Wireless Solar Water Splitting Using Silicon-Based Semiconductors and Earth-Abundant Catalysts," *Science,* vol. 334, pp. 645-648, 2011.

[3] J. Luo, J.-H. Im, M. T. Mayer, M. Schreier, M. K. Nazeeruddin, N.-G. Park, S. D. Tilley, H. J. Fan and M. Grätzel, "Water photolysis at 12.3% efficiency via perovskite photovoltaics and Earth-abundant catalysts," *Science,* vol. 345, no. 6204, pp. 1593-1596, 2014.

[4] P. Bogdanoff, D. Stellmach, O. Gabriel, B. Stannowski, R. Schlatmann, R. van de Krol and S. Fiechter, "Artificial Leaf for Water Splitting Based on a Triple-Junction Thin-Film Silicon Solar Cell and a PEDOT:PSS/Catalyst Blend," *Energy Technology,* vol. 4, pp. 230-241, 2016.

[5] L. Schlapbach and A. Züttel, "Hydrogen-storage materials for mobile applications," *Nature,* vol. 414, p. 353–358, 2001.

[6] U. Prasad, J. Prakash, B. Azeredo and A. Kannan, "Stoichiometric and non-stoichiometric tungsten doping effect in bismuth vanadate based photoactive material for photoelectrochemical water splitting," *Electrochimica Acta,* vol. 299, pp. 262-272, 2019.

[7] A. Fujishima and K. Honda, "Electrochemical Photolysis of Water at a Semiconductor Electrode," *Nature,* pp. 37-38, 1972.

[8] O. Khaselev and J. A. Turner, "A Monolithic Photovoltaic-Photoelectrochemical Device for Hydrogen Production Via Water Splitting," *Science,* vol. 280, pp. 425-427, 1998.

[9] B. Weng, Z. Xiao, W. Meng, C. R. Grice, T. Poudel, X. Deng and Y. Yan, "Bandgap Engineering of Barium Bismuth Niobate Double," *Advanced Energy materials,* 2017.

[10] J. Gu, Q. Huang, Y. Yuan, K.-H. Ye, Z. Wang and W. Mai, "In situ growth of a TiO2 layer on a flexible Ti substrate targeting the interface recombination issue of BiVO4 photoanodes for efficient solar water splitting," *Journal of Materials Chemistry A,* 2017.

[11] M. Rohloff, B. Anke, S. Zhang, U. Gernert, C. Scheu, M. Lerch and A. Fischer, "Mo-doped BiVO4 thin films – high photoelectrochemical water splitting performance achieved by a tailored structure and morphology," *Sustaible Energy & Fuels,* pp. 1830-1846, 2017.

[12] H. D. Telpande and D. V. Parwate, "Characterization supported improved method for the synthesis of bismuth vanadate and its assessment with conventional synthetic route," *Journal of Applied Chemistry,* vol. 8, no. 5, pp. 28-37, 2015.

[13] A.-d. l. Cruz and U. GarcíaPérez, "Photocatalytic properties of BiVO4 prepared by the co-precipitation method: Degradation of rhodamine B and possible reaction mechanisms under visible irradiation," *Materials Research Bulletin,* vol. 45, no. 2, pp. 135-141, 2010.

[14] L. Zhang, D. Chen and X. Jiao, "Monoclinic Structured BiVO4 Nanosheets: Hydrothermal Preparation, Formation Mechanism, and Coloristic and Photocatalytic Properties," *Journal of Physical Chemistry B,* vol. 110, pp. 2668-2673, 2006.

[15] Y. Lin, C. Lu and C. Wei, "Microstructure and photocatalytic performance of BiVO4 prepared by hydrothermal method," *Journal of Alloys and Compounds,* vol. 781, pp. 56-63, 2019.

[16] J. Yu and A. Kudo, "Effects of Structural Variation on the Photocatalytic Performance of Hydrothermally Synthesized BiVO4," *Advanced Functional Materials,* vol. 16, no. 16, pp. 2163-2169, 2006.

[17] S. Tokunaga, H. Kato and A. Kudo, "Selective Preparation of Monoclinic and Tetragonal BiVO4 with Scheelite Structure and Their Photocatalytic Properties," *Chemistry of materials,* vol. 13, no. 12, pp. 4624-4628, 2001.

[18] Z. Qu, P. Liu, X. Yang, F. Wang, W. Zhang and C. Fei, "Microstructure and Characteristic of BiVO4 Prepared under Different pH Values: Photocatalytic Efficiency and Antibacterial Activity," *Materials (Basel),* vol. 9, 2016.

[19] K. T. Butler, B. J. Dringoli, L. Zhou, P. M. Rao, A. Walsh and L. V. Titova, "Ultrafast carrier dynamics in BiVO4 thin film photoanode material: interplay between free carriers, trapped carriers and low-frequency lattice vibrations," *Journal of Materials Chemistry A,* vol. 4, pp. 18516-18523, 2016.

[20] T. W. Kim and K.-S. Choi, "Nanoporous BiVO4 Photoanodes with Dual-Layer Oxygen Evolution Catalysts for Solar Water Splitting," *Science,* vol. 343, pp. 990-994, 2014.

[21] F. F. Abdi, T. J. Savenije, M. M. May, B. Dam and R. v. d. Krol, "The Origin of Slow Carrier Transport in BiVO4 Thin Film Photoanodes: A Time-Resolved Microwave Conductivity Study," *The Journal of Physical Chemistry Letters,* vol. 4, no. 16, pp. 2752-2757, 2013.

[22] D. Zhong, S. Choi and D. Gamelin, "Near-complete suppression of surface recombination in solar photoelectrolysis by "Co-Pi" catalyst-modified W:BiVO4.," *Journal of American Chemical Society,* vol. 133, pp. 18370-18377, 2011.

[23] F. F. Abdi, N. Firet and R. van de Krol, "Efficient BiVO4 Thin Film Photoanodes Modified with Cobalt Phosphate Catalyst and W-doping," *ChemCatChem,* vol. 5, no. 2, pp. 490-496, 2013.

[24] J.-M. Wu, Y. Chen, L. Pan, P. Wang, Y. Cui, D. Kong, L. Wang, X. Zhang and J.-J. Zou, "Multi-layer monoclinic BiVO4 with oxygen vacancies and V4+ species for highly efficient visible-light photoelectrochemical applications," *Applied Catalysis B: Environmental,* vol. 221, pp. 187-195, 2018.

[25] S. K. Pilli, T. E. Furtak, L. D. Brown, T. G. Deutsch, J. A. Turner and A. M. Herring, "Cobalt-phosphate (Co-Pi) catalyst modified Mo-doped BiVO4 photoelectrodes for solar water oxidation," *Energy & Environment Science,* vol. 4, no. 12, pp. 5028-5034, 2011.

[26] M. Huang, J. Bian, W. Xiong, C. Huang and R. Zhang, "Low-dimensional Mo:BiVO4 photoanodes for enhanced photoelectrochemical activity," *Journal of Materials Chemistry A,* vol. 6, no. 8, pp. 3602-3609, 2018.

[27] J. Quiñonero and R. Gómez, "Controlling the amount of co-catalyst as a critical factor in determining the efficiency of photoelectrodes: The case of nickel (II) hydroxide on vanadate photoanodes," *Applied Catalysis B: Environmental,* vol. 217, pp. 437-447, 217.

[28] D. Kong, J. Qi, D. Liu, X. Zhang, L. Pan and J. Zou, "Ni-Doped BiVO4 with V4+ Species and Oxygen Vacancies for Efficient Photoelectrochemical Water Splitting," 2019.

[29] H. P. Sarker, P. M. Rao and M. N. Huda, "Niobium Doping in BiVO4: Interplay Between Effective Mass, Stability, and Pressure," *Chemical Physical Chemistry,* vol. 20, pp. 773-784, 2019.

[30] A. F. Fatwa, H. Lihao, A. H. M. Smets, M. Zeman, D. Bernard and R. V. d. Krol, "Efficient solar water splitting by enhanced charge separation in a bismuth vanadate-silicon tandem photoelectrode," *Nature Communicataions,* 2013.

[31] B. J. Trześniewski and W. A. Smith, "Photocharged BiVO4 photoanodes for improved solar water splitting," *Journal of Materials Chemistry,* vol. 4, no. 8, pp. 2919-2926, 2016.

[32] A. Loiudice, J. Ma, W. S. Drisdell, T. M. Mattox, J. K. Cooper, T. Thao, C. Giannini, J. Yano, L.-W. Wang, I. D. Sharp and R. Buonsanti, "Bandgap Tunability in Sb-alloyed BiVO4 Quaternary Oxides as Visible Light Absorbers for Solar fuel Applications," *Advanced Materials,* pp. 6733-6740, 2015.

[33] Y. Park, K. J. McDonald and K.-S. Choi, "Progress in bismuth vanadate photoanodes for use in solar water oxidation," *Chemical Society Reviews,* vol. 42, pp. 2321-2337, 2013.

[34] D. K. Lee, D. Lee, M. A. Lumley and K.-S. Choi, "Progress on ternary oxide-based photoanodes for use in photoelectrochemical cells for solar water splitting," *Chemical Society Reviews,* vol. 48, pp. 2126-2157, 2019.

[35] R. L. Doyle, I. J. Godwin, M. P. Brandon and M. E. G. Lyons, "Redox and electrochemical water splitting catalytic properties of hydrated metal oxide modified electrodes," *Physical Chemistry Chemical Physics,* vol. 15, no. 13, pp. 13737-13783 , 2013.

[36] K. U. Wijayantha, S. Saremi-Yarahmadi and L. M. Peter, "Kinetics of oxygen evolution at α-Fe2O3 photoanodes: a study by photoelectrochemical impedance spectroscopy," *Physical Chemistry Chemical Physics,* vol. 13, pp. 5264-5270, 2011.

[37] L. Zhou, C. Zhao, B. Giri, P. Allen, X. Xu, H. Joshi, Y. Fan, L. V. Titova and P. M. Rao, "High Light Absorption and Charge Separation Efficiency at Low Applied Voltage from Sb-Doped SnO2/BiVO4 Core/Shell NanorodArray Photoanodes," *Nano Letters,* vol. 16, p. 3463−3474, 2016.

[38] L. Zhou, Y. Yang, J. Zhang and P. M. Rao, "Photoanode with Enhanced Performance Achieved by Coating BiVO4 onto ZnO-Templated Sb-Doped SnO2 Nanotube Scaffold," *Applied Materials & Interfaces,* vol. 9, pp. 11356-11362, 2017.

[39] M. G. Lee, D. H. Kim, W. Sohn, C. W. Moon, H. Park, S. Lee and H. W. Jang, "Conformally coated BiVO4 nanodots on porosity-controlled WO3 nanorods as highly efficient type II heterojunction photoanodes for water oxidation," *Nano Energy,* vol. 28, pp. 250-260, 2016.

[40] D. Kang, Y. Park, J. C. Hill and K.-S. Choi, "Preparation of Bi-Based Ternary Oxide Photoanodes BiVO4, Bi2WO6, and Bi2Mo3O12 Using Dendritic Bi Metal Electrodes," *The Journal of Physical Chemistry Letters,* vol. 5, pp. 2994-2999, 2014.

[41] A. Tayyebi, T. Soltani and B.-K. Lee, "Effect of pH on photocatalytic and photoelectrochemical (PEC) properties of monoclinic bismuth vanadate," *Journal of Colloid and Interface Science,* vol. 534, pp. 37-46, 2019.

[42] M. Tayebi and B.-K. Lee, "Recent advances in BiVO4 semiconductor materials for hydrogen production using photoelectrochemical water splitting," *Renewable and Sustainable Energy Reviews,* vol. 111, pp. 332-343, 2019.

[43] D. Lee, A. Kvit and K.-S. Choi, "Enabling Solar Water Oxidation by BiVO4 Photoanodes in Basic Media," *Chemistry of Materials,* vol. 30, p. 4704−4712, 2018.

[44] M. Tayebi and B.-K. Lee, "Recent advances in BiVO4 semiconductor materials for hydrogen production using photoelectrochemical water splitting," *Renewable and Sustainable Energy Reviews,* vol. 211, pp. 332-343, 2019.

[45] G. Segev, H. Dotan, K. D. M. A. Kay, M. T. Mayer, M. Grätzel and A. Rothschild, "High Solar Flux Concentration Water Splitting with Hematite (α-Fe 2 O 3 ) Photoanodes," *Advanced Energy Materials,* 2015.

[46] J. Y. Kim, G. Magesh, D. H. Youn, J.-W. Jang, J. Kubota, K. Domen and J. S. Lee, "Single-crystalline, wormlike hematite photoanodes for efficient solar water splitting," *Scientific Reports,* vol. 3, p. 2681, 2013.

[47] G. V. Govindaraju, J. M. Morbec, G. A. Galli and K.-S. Choi, "Experimental and Computational Investigation of Lanthanide Ion Doping on BiVO4 Photoanodes for Solar Water Splitting," *The Journal of Physical Chemistry,* vol. 122, pp. 19416-19424, 2018.